%% file: h22vm2ve.tex
\documentclass[11pt, a4paper]{article}
\pdfoutput=1
\usepackage{jheppub}
\usepackage{amsmath}
\usepackage{graphicx}
\usepackage[config=altsf]{subfig}
\usepackage{feynmf}
\usepackage{dsfont}
\usepackage{array}
\usepackage{slashed}
\usepackage{afterpage}
\usepackage{appendix}
\usepackage{multirow}
\usepackage{dcolumn}
\usepackage{slashed}
\usepackage{soul}

\usepackage{booktabs}    
\usepackage{placeins}
\usepackage{bbm}
\usepackage{amsfonts}
\usepackage{amsmath,amssymb}
\usepackage{mathrsfs}
\usepackage{epsfig}
\usepackage{graphicx}
\usepackage{wrapfig}                 
\usepackage{url}
\usepackage{hyperref}
\usepackage{float}
\usepackage{pstricks}
\usepackage{color}
\usepackage{multirow}
\usepackage{lipsum}
\usepackage{enumitem}
\usepackage{ctable}
\newcolumntype{L}{>{\centering\arraybackslash}m{1.5cm}}

\usepackage{pgfkeys, pgfsys, pgfcalendar}

\newcommand{\bear}{\begin{array}}
\newcommand {\eear}{\end{array}}

\newcommand{\beq}{\begin {equation}}  
\newcommand{\eeq}{\end   {equation}} 
\newcommand{\bea}{\begin {eqnarray}} 
\newcommand{\eea}{\end   {eqnarray}}  
\newcommand{\beqa}{\begin {eqnarray}} 
\newcommand{\eeqa}{\end   {eqnarray}}  
\newcommand{\baa}{\begin {array}   } 
\newcommand{\eaa}{\end   {array}   }     
\newcommand{\bit}{\begin {itemize} }
\newcommand{\eit}{\end   {itemize} }
\newcommand{\be }{\begin {equation}} 
\newcommand{\ee }{\end   {equation}}

\newcommand{\bef}{\begin{figure}}
\newcommand {\eef}{\end{figure}}
\newcommand{\bec}{\begin{center}}
\newcommand {\eec}{\end{center}}

\definecolor{cerulean}{rgb}{0., 0.62,0.7}

\newcommand{\twiddle}{{\raise.17ex\hbox{$\scriptstyle\sim$}}}

\vspace*{3cm}

\title{\bf Effect of anomalous $HHH$ and $ZZHH$ couplings on the decay width of $H\rightarrow \nu_e\bar{\nu}_e\nu_\mu\bar{\nu}_\mu$ }

\author[a,b,c]{Pankaj Agrawal} 
\author[a,b,d]{and Biswajit Das}

\affiliation[a]{Institute of Physics, \\Sainik School Post, Bhubaneswar 751 005, India}
\affiliation[b]{Homi Bhabha National Institute,\\Training School Complex, Anushakti Nagar, Mumbai 400094, India}
\affiliation[c]{Center for Quantum Engineering, Research and Education (CQuERE), TCG CREST, Salt Lake Sector 5, Kolkata 700091, India}
\affiliation[d]{The Institute of Mathematical Sciences,
IV Cross Road, Taramani, Chennai 600113, India}

\emailAdd{agrawal@iopb.res.in}
\emailAdd{bisubangla92@gmail.com}

\abstract{ Despite the discovery of the Higgs boson,
  the Higgs sector of the standard model is still not fully established.
  In particular, the self couplings of the Higgs boson, and its
  couplings with gauge bosons, are still to be fully determined.
  We consider eletroweak corrections to the process 
  $H\rightarrow \nu_e\bar{\nu}_e\nu_\mu\bar{\nu}_\mu$. The 
  corrections depend on the $HHH$ and $ZZHH$ couplings.
  We investigate this dependence in $\kappa$-framework.
  We find that the width depends on $HHH$ coupling significantly.
  The dependence on $ZZHH$ coupling is only marginal. We also
  discuss the dependence on $ZZWW$ coupling.  }

\begin{document}
\maketitle

\flushbottom

\newpage

\section{Introduction}

\label{sec:intro}

   For about fifty years, Standard Model has been tested at many
 colliders, including at the Large Hadron Collider (LHC). No significant departure
 from the model predictions have been found \cite{conference1,conference2}. There are some open questions, like the nature of neutrino mass, dark matter,
 baryogenesis, etc, that may necessitate going beyond the standard model. But there is no hint in the observational and experimental 
 data for any specific model beyond the standard model. In 2012, the last missing piece of the standard model, the Higgs boson, was detected at
 the LHC. Since then, the properties of this particle have been 
 studied. These properties are consistent with the standard
 model. However, some of the couplings of the model are still 
 to be determined fully. This has left open the question of
 the shape of the Higgs potential  \cite{Agrawal:2019bpm}. It is important to measure
 all the properties of the Higgs boson with good enough precision
 to demonstrate the complete validity of the standard model.
 
 The couplings of the Higgs boson can be determined either
 through a production process, or the decays. Some of the
 couplings of the Higgs boson, like its self couplings, and
 quartic couplings with gauge bosons are hard to determine.
 Even at the high-luminosity LHC (HL-LHC), it is not clear
 if these couplings can be measured with good enough precision.
 This is because the requirement of multiple Higgs bosons
 and/or gauge bosons
 in the production process.
 Another avenue to determine these couplings is electroweak
 radiative corrections to either a single Higgs boson production
 process (like vector-boson fusion or associated production
 with a vector boson) or decays like $H \to V V^*$ (V is either
 W or Z boson.). One can also study two loop corrections to
 a few processes to explore possibility of measuring
 these couplings.

  In this Letter, we consider the decay process $H\rightarrow \nu_e\bar{\nu}_e\nu_\mu\bar{\nu}_\mu$. We compute one-loop electroweak corrections to this process. These corrections depend on the trilinear Higgs boson
  coupling, $HHH$, and the quartic coupling $ZZHH$. We explore the possibility of measuring these couplings in this decay process. 
  We use $\kappa$-framework to determine this dependence. We have focused initially
  at this process to avoid complications due to final state radiation
  when there is a charged lepton in the final state. That will be done
  at a later stage. There are a few groups who have been studied electroweak corrections to $H\rightarrow 4l$ channel \cite{BREDENSTEIN2006131,Boselli:2015aha}.  There are loose experimental bounds on these couplings.  Using 126 fb$^{-1}$ of data at 13
    TeV, the ATLAS collaboration has put a bound on the couplings using the production of
 a pair of Higgs boson and subsequent decay into $b{\bar b}b{\bar b}$. They obtained  bounds of $ -5.4 < \kappa_{HHH} < 11.4$ and $ -0.10 < \kappa_{V_2H_2} < 2.56$
 at 95$\%$ confidence level \cite{Aad_2021, atlascollaboration2023search}. Here $ \kappa_{V_2H_2}$ is the scaling factor for the $VVHH$ coupling and $ \kappa_{HHH}$ is the scaling factor for the $HHH$ coupling. However, in this process both coupligs $WWHH$ and $ZZHH$ are present. The
    process $p p \to HHV$, with a $W$ or a $Z$ boson, allows us to separately measure $WWHH$ and $ZZHH$ couplings. The expected bound
    from the $WHH$ production
  at the HL-LHC is $ -9.4 < \kappa_{V_2H_2} < 7.9$ \cite{Nordstrom:2018ceg}, which is quite weak. The decay process under consideration
  depends on $ZZHH$, not on $WWHH$.
 According to the future projections for HL-LHC, $-0.5 < \kappa_{HHH} < 6.1$ and
$0.7 < \kappa_{V_2H_2} < 1.4$ at 68$\%$ confidence level \cite{ATLAS:2022faz}.
 
   The paper is organized as follows. In the next section we discuss
   the process and the diagrams that contribute to it. In the section 3, we discuss how we did the calculation. In the section 4, we discuss 
   our results. In the last section, we have some conclusions.

\section{The Process}
\label{sec:prcs}
We are interested in calculating the decay width of the decay channel $H\rightarrow \nu_e \bar{\nu}_e \nu_\mu \bar{\nu}_\mu$. As our aim to see the effect of anomalous $HHH$ and $ZZHH$ couplings in this channel, we calculate NLO electroweak (EW) correction. There are a few one-loop diagrams where we can vary $HHH$ and $ZZHH$ couplings to study the effect on the decay width.

\begin{figure}[!h]
  \begin{center}
\includegraphics [angle=0,width=0.5\linewidth]{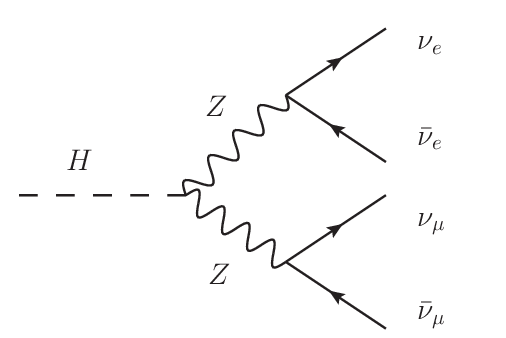}\\
	\caption{LO Feynman diagram for decay $H\rightarrow \nu_e \bar{\nu}_e \nu_\mu \bar{\nu}_\mu$.  }
	\label{fig:tree_dia}
	\end{center}
\end{figure}

\begin{figure}[!h]
  \begin{center}
\includegraphics [angle=0,width=1\linewidth]{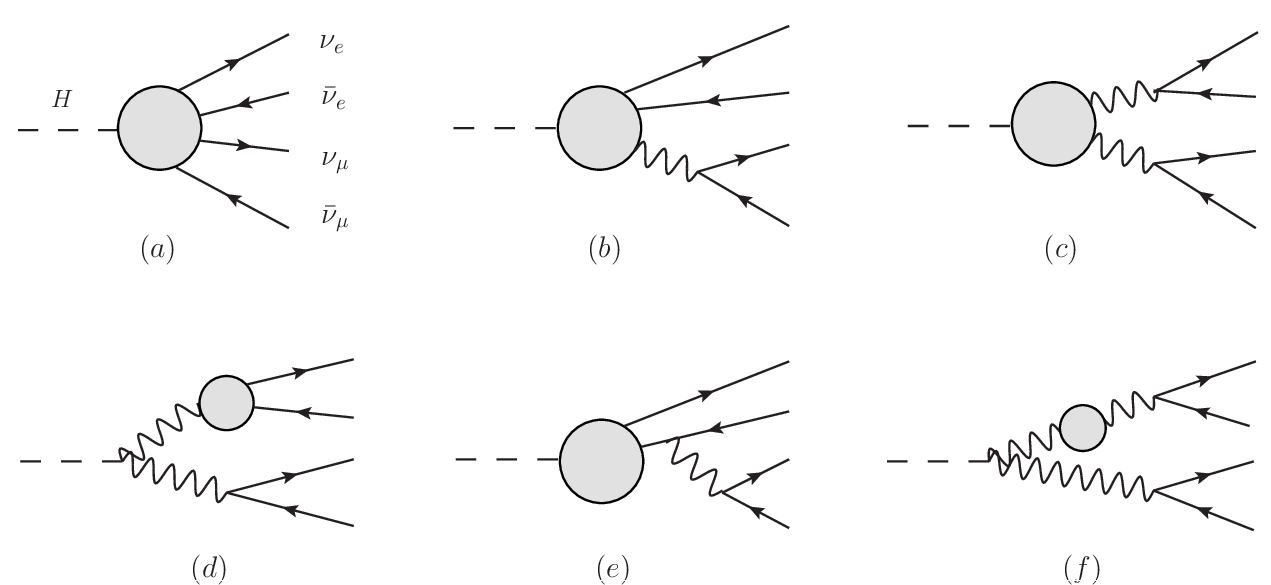}\\
	\caption{Generic NLO EW virtual Feynman diagrams for $H\rightarrow \nu_e \bar{\nu}_e \nu_\mu \bar{\nu}_\mu$.  }
	\label{fig:loop_dia}
	\end{center}
\end{figure}

At the leading order (LO), there is only one Feynman diagram as shown in Fig.~\ref{fig:tree_dia}, as we allow one $Z$ to decay into muon  neutrinos and another into electron neutrinos. In the calculation of
the one-loop (OL) level amplitudes, there are a total of $118$ Feynman diagrams. As the process is $1\rightarrow 4$, the virtual diagrams are of pentagonal, box, triangle and bubble type. The generic one-loop diagrams are shown in Fig.~\ref{fig:loop_dia}.

\begin{figure}[!h]
  \begin{center}
\includegraphics [angle=0,width=1\linewidth]{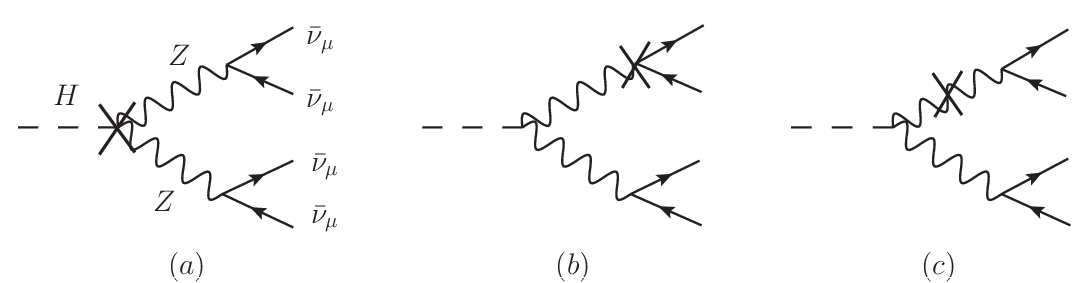}\\
	\caption{Counterterm diagrams for NLO EW correction to $H\rightarrow \nu_e \bar{\nu}_e \nu_\mu \bar{\nu}_\mu$.  }
	\label{fig:ct_dia}
	\end{center}
\end{figure}
\begin{figure}[!h]
  \begin{center}
\includegraphics [angle=0,width=1\linewidth]{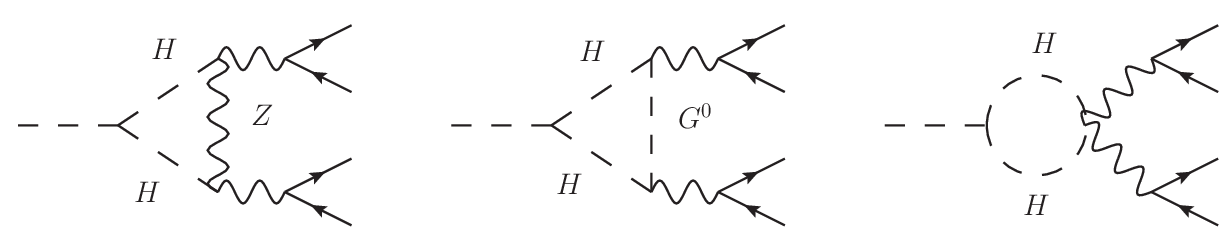}\\
	\caption{ NLO EW virtual diagrams with $HHH$ and $ZZHH$ couplings.  }
	\label{fig:hhh_dia}
	\end{center}
\end{figure}

\begin{figure}[!h]
  \begin{center}
\includegraphics [angle=0,width=1\linewidth]{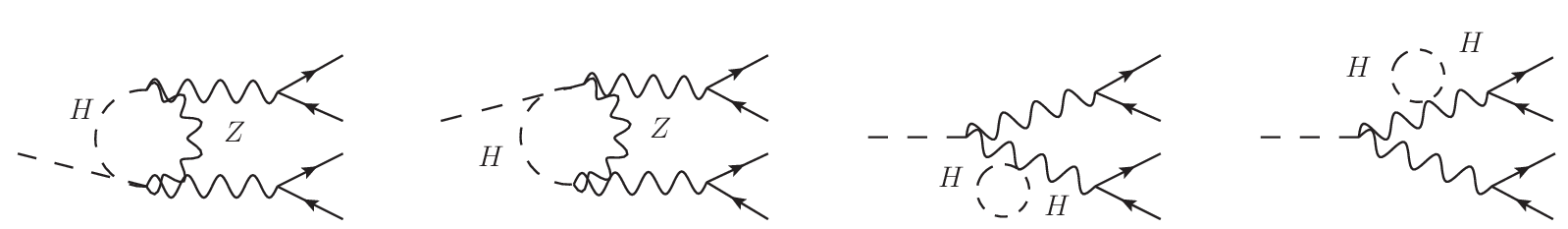}\\
	\caption{ NLO EW virtual diagrams with $ZZHH$ couplings.  }
	\label{fig:zzhh_dia}
	\end{center}
\end{figure}
\begin{figure}[!h]
  \begin{center}
\includegraphics [angle=0,width=1\linewidth]{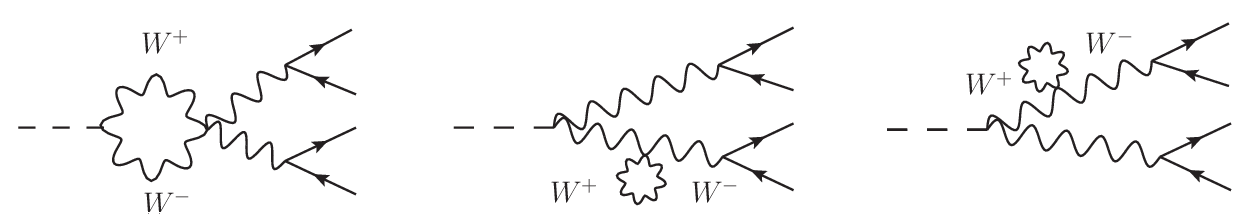}\\
	\caption{ NLO EW virtual diagrams with $ZZWW$ couplings.  }
	\label{fig:wwzz_dia}
	\end{center}
\end{figure}

The generic diagram $(a)$ displayed in Fig.~\ref{fig:loop_dia}, is a pentagon type diagram. There are $6$ pentagon diagrams in this process.  The generic diagram $(b)$ displayed in Fig.~\ref{fig:loop_dia}, is a box type diagram. There is a another similar generic diagram with other neutrinos that is not shown in Fig.~\ref{fig:loop_dia}. There are $12$ box type diagrams in this process. The generic $(c)$ diagram is the correction to $HZZ$-vertex. This generic diagram has both triangle and bubble type diagrams. We have total $40$ virtual diagrams related to this generic diagram, out of which $30$ are triangle and $10$ are bubble type diagrams. $HZZ$-vertex correction diagrams are relevant in this study, as there are a few diagrams where we can introduce the effect of anomalous $HHH$, $ZZHH$ and $ZZWW$ couplings. The related diagrams are shown in Fig.~\ref{fig:hhh_dia}, Fig.~\ref{fig:zzhh_dia} and, Fig.~\ref{fig:wwzz_dia}. The generic diagram $(d)$ in the second row of Fig.~\ref{fig:loop_dia}, is the virtual correction to $Z\nu\bar{\nu}$ vertex. We also have one more similar generic diagram for other $Z\nu\bar{\nu}$ vertex. There is a total of $6$ triangle type virtual diagrams related to these generic diagrams. The generic diagram $(e)$ in the second row of Fig.~\ref{fig:loop_dia}, is the triangle type diagram and there is a another similar generic diagram with Higgs and other neutrinos. There are $8$ triangle type diagrams related to these generic diagrams. The last generic diagram $(f)$ in Fig.~\ref{fig:loop_dia}, represents the $Z$-boson self energy diagrams. There is another similar generic diagram with another $Z$-boson that has not been shown in Fig.~\ref{fig:loop_dia}. All such diagrams are bubble type diagrams. There is also dependence on $ZZHH$ and $ZZWW$ coupling from the $Z$-boson self energy diagrams. The related diagrams are shown in Fig.~\ref{fig:zzhh_dia} and Fig.~\ref{fig:wwzz_dia}. We have also listed the counterterm (CT) diagrams in Fig.~\ref{fig:ct_dia}. There are another two CT diagrams similar to diagram $(b)$ and $(c)$ related to other $Z\nu\bar{\nu}$-vertex and other $Z$-boson self energy diagrams. These five CT diagrams cancel all UV divergence from the virtual amplitudes. As shown in the Fig.~\ref{fig:ct_dia}, diagram $(a)$ is the CT diagram for $HZZ$-vertex, diagram $(b)$ is the CT diagram for $Z\nu\bar{\nu}$-vertex and, diagram $(c)$ is the CT diagram for $Z$-self energy diagrams. The computation of CT diagrams also involve the self-energy 
diagrams corresponding to $Z$-boson, $W$-boson, and, Higgs boson where we can introduce anomalous  $HHH$, $ZZHH$ and, $ZZWW$ couplings. A detail study about the effect of anomalous coupling has been discussed in Sec.~\ref{subsec:renorm_cms}. There are no real emission diagrams as the tree level diagram(Fig.~\ref{fig:tree_dia}) do not have any charged gauge bosons or charged leptons to emit photons. All diagrams have been generated using {\tt FeynArts} \cite{Hahn:2000kx}, a {\tt Mathematica} package.

\section{Calculations}
\label{sec:calc_check}
There are a few hundreds diagrams at one-loop level. As we treat leptons and light quarks to be massless, a large set of diagrams become zero because of vanishing coupling with the scalars. We also ignore the tadpole diagrams as the renormalization condition will set them to zero. With these considerations, we are left with $118$ diagrams to compute as mentioned in Sec.~\ref{sec:prcs}. There are another type of diagrams similar to generic diagram $(f)$ in Fig.~\ref{fig:loop_dia}, which are the bubble type diagrams for the
mixed propagators between a Goldstone boson and $Z$-boson. These diagrams do not contribute.  We use helicity formalism to compute the process amplitudes at one-loop level as well as at tree level. First we classify the one-loop level diagrams in a few set of prototype amplitudes. Then we compute all virtual diagrams with the help of these prototype amplitudes by suitable crossing, mass and coupling choices. Tree level helicity amplitudes can be computed easily with the spinor products $[pq]$ and $\langle pq \rangle$ \cite{Peskin:2011in}. To calculate one-loop level helicity amplitude, we also use a vector current $\langle p \gamma^\mu q ]$ \citep{Agrawal:2021owa}. We calculate one-loop amplitudes in 't Hooft and Veltman Scheme(HV) scheme \citep{THOOFT1972189}. In this scheme the loop part('observed') is computed in $d$-dimension and rest of the amplitude('unobserved') is computed in $4$-dimension \citep{PhysRevD.84.094021}. The gamma matrix, loop momentum algebra in {{'observed'}} part has been done in $d$-dimension. In this process, we have two fermion-loop ($t$-quark) diagrams where we face $\gamma^5$-anomaly issues. The detailed discussion on $\gamma^5$-anomaly is in the next sub-section~(\ref{subsec:gamma_five_anomaly}).

We use the symbolic manipulation program {\tt FORM} \cite{Vermaseren:2000nd}, to calculate the helicity amplitudes. Using {\tt FORM}, the amplitudes are written in terms of spinor products, scalar product of momenta and different vector objects. One needs to also calculate one-loop scalar and tensor integral for NLO amplitude computation. The one-loop scalar integrals have been calculated using a package {\tt OneLoop} \citep{vanHameren:2010cp}. To calculate the tensor integral, we use an in-house reduction code, OVReduce \cite{Agrawal:2012df,Agrawal:1998ch}. At last, the phase space integral has been computed with the {{Monte Carlo
integration package {\tt AMCI}~\cite{Veseli:1997hr}. In the AMCI package, the {\tt VEGAS} algorithm \cite{Lepage:1977sw} has been implemented using parallel
virtual machine ({\tt PVM}) package \cite{10.7551/mitpress/5712.001.0001}.}}
\subsection{$\gamma^5$-anomaly}
\label{subsec:gamma_five_anomaly}
\begin{figure}[!h]
  \begin{center}
\includegraphics [angle=0,width=0.8\linewidth]{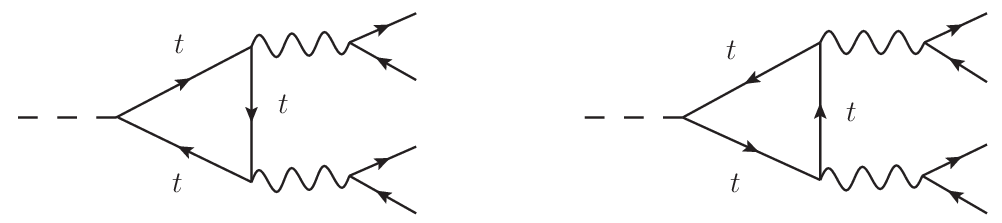}\\
	\caption{ The triangle virtual diagrams with $t$-quark fermion loop.}
	\label{fig:top_dia}
	\end{center}
\end{figure}

At one-loop level, we have two triangle diagrams involving $t$-quark fermion loop. The corresponding diagrams are shown in Fig.~\ref{fig:top_dia}. To find these amplitudes, one needs to compute trace involving $\gamma^5$-matrices that come from two $Zt\bar{t}$ vertices. This trace is inconsistent, as one can get different results depending on the different starting points of the trace. The formal definition of $\gamma^5=\frac{i}{4!}\epsilon_{\mu\nu\rho\sigma}\gamma^\mu\gamma^\nu\gamma^\rho\gamma^\sigma$ is not consistent in $d$-dimension as the anti-symmetric tensor $\epsilon_{\mu\nu\rho\sigma}$ lives in $4$-dimension, so, $\gamma^5$ do not anti-commute with other $\gamma$-matrices in $d$-dimension. The problem arises because of simultaneous use of cyclic properties of the trace and anti-commutation relation between $\gamma^5$ and other $\gamma$-matrices. Therefore, one of the properties needs to be dropped to get the right result
\citep{PhysRevD.84.094021}. 

To address this issue, two elegant treatments have been introduced so far. One is known as BMHV (Breitenlohner-Maison-'t Hooft-Veltman) scheme and other one is known as KKS (Korner-Kreimer-Schilcher) scheme \citep{PhysRevD.84.094021,Korner:1991sx}. In our calculation, we use KKS scheme to calculate these traces shown in Fig.~\ref{fig:top_dia}. Following KKS prescription, we take all $\gamma^5$-matrices to a particular vertex('reading point') by anti-commuting with other $\gamma$-matrices. Then we do $d$-dimensional algebra and compute trace. We also follow same prescription to calculate $Z\nu\bar{\nu}$-vertex correction where we have a current with $d$-dimensional $\gamma$-matrices and  $\gamma^5$-matrices \citep{Garzelli:2009is}. This removes $\gamma^5$-anomaly from our calculation.
\subsection{Renormalization and CMS at one loop}
\label{subsec:renorm_cms}
We use on-shell renormalization scheme to calculate the required CTs for this process. Following the definition and notation of Ref. \cite{Denner:1991kt}, we write the required bare parameters in terms of renormalized parameters and corresponding counterterms as follows

\begin{eqnarray}
e_0=(1+\delta Z_e)e,\:\: \:M_{W,0}^2=M^2_W+\delta M_W^2, \quad{\text {and}}\quad M_{Z,0}^2=M^2_Z+\delta M_Z^2.
\label{eq:renor_papra}
\end{eqnarray}
We also need wave function renormalization constants of the fields that are involved in this process to remove the UV divergences. The required wave function renormalization constants are defined as   
\begin{eqnarray}
Z_0=(1+\frac{1}{2}\delta Z_{ZZ})\; Z+\frac{1}{2}\delta Z_{ZA}\; A,\quad A_0=(1+\frac{1}{2}\delta Z_{AA})\; A+\frac{1}{2}\delta Z_{AZ}\; Z,\nonumber \\
W_0^\pm=(1+\frac{1}{2}\delta Z_W)\; W^\pm,\:\:\:H_0=(1+\frac{1}{2}\delta Z_H)\; H,\:\;\: f_{i,0}^L=(\delta_{ij}+\frac{1}{2}\delta Z_{ij}^{f,L})\; f_j^L.
\label{eq:renor_wf}
\end{eqnarray}
The bare quantities are denoted by subscript $0$. We take diagonal CKM matrix ($V_{CKM}$) and no renormalization is needed for $V_{CKM}$. We also take all leptons and light fermions to be massless.  The counterterms can be calculated  by imposing suitable renormalization conditions as given in Ref.\cite{Denner:1991kt}. The required renormalization constants are:
\begin{gather}
\delta M_W^2={\text {Re}}\:\Sigma ^W_T(M_W^2),\quad \delta Z_W=-{\text{Re}}\frac{\partial}{\partial k^2}\Sigma^W_T(k^2)\bigg |_{k^2=M_W^2},\nonumber\\
\delta M_Z^2={\text {Re}}\:\Sigma ^{ZZ}_T(M_Z^2),\quad \delta Z_{ZZ}=-{\text{Re}}\frac{\partial}{\partial k^2}\Sigma^{ZZ}_T(k^2)\bigg |_{k^2=M_Z^2},\nonumber\\
\delta Z_{ZA}=2\frac{\Sigma^{AZ}_T(0)}{M_Z^2},\quad \delta Z_{AA}=-\frac{\partial}{\partial k^2}\Sigma^{AA}_T(k^2)\bigg |_{k^2=0},\quad\nonumber\\
\delta Z_e=-\frac{1}{2}\delta Z_{AA}-\frac{s_W}{c_W}\frac{1}{2}\delta Z_{ZA},\quad\delta Z_{H}=-{\text{Re}}\frac{\partial}{\partial k^2}\Sigma^H(k^2)\bigg |_{k^2=M_H^2},\nonumber\\
\delta Z_{ii}^{f,L}=-{\text {Re}}\:\Sigma_{ii}^{f,L}(m_{f,i}^2)-m_{f,i}^2\frac{\partial}{\partial k^2}{\text {Re}}[\Sigma_{ii}^{f,L}(k^2)+\Sigma_{ii}^{f,R}(k^2)+2\Sigma_{ii}^{f,S}(k^2)]\bigg |_{k^2=m_{f,i}^2}.
\label{eq:ct_terms_cal}
\end{gather}
The self energies $\Sigma$, given in Eq.~\ref{eq:ct_terms_cal} can be calculated from self energy of $W$-boson, $Z$-boson, photon (also $Z\gamma$) and $H$-boson propagators. We have neglected tilde($\sim$) symbol from 'Re' as we consider real diagonal CKM matrix. We have calculated these $\Sigma$ and then counterterms given in  Eq.~\ref{eq:ct_terms_cal} in the HV scheme. We find exactly same results as given in Ref. \cite{Denner:1991kt}. The self energy diagrams of $Z$-boson have $ZZHH$ and $ZZWW$ coupling dependencies. Similarly, $H$-self energy diagrams can have $HHH$ coupling and $W$-boson self-energy diagrams can have $ZZWW$ coupling dependencies. As our primary goal is to study the effect of these anomalous couplings, we also need to scale these couplings in CT diagrams. The Feynman rules for counterterm diagrams given in Fig.~\ref{fig:ct_dia} have been taken from Ref. \cite{Denner:1991kt}. With the counterterms, all UV pole($\frac{1}{\epsilon}$) cancel from the one-loop virtual diagrams. 

We use complex mass scheme(CMS) \cite{Denner:2006ic} to treat unstable particle in one-loop electroweak corrections. In CMS, the unstable masses are defined with a complex part as 
\begin{eqnarray}
m_V^2\rightarrow\mu_V^2=m_V^2-im_V\Gamma_V,
\label{eq:cms_equ_cal}
\end{eqnarray} 
where $V=W,Z$ and $\Gamma_V$ is the corresponding decay width. This treatment also makes Weinberg angle complex as $\cos^2\theta_W=\mu^2_W/\mu^2_Z$. For the $t$-quarks, same prescription has been followed. These complex masses and $\cos\theta$ have been used everywhere in the perturbative calculation to maintain the gauge invariance. The renormalization in CMS has been done in a modified version of the on-shell scheme \cite{Denner:2006ic,Denner:2005fg}. In this treatment, the renormalized mass is the pole of the corresponding propagator in the complex plane. When renormalized conditions are imposed, one needs to perform the self-energy computation with complex momenta. This computation can be done with Taylor expansion of self energies about the real mass and maintaining the one loop accuracy. We follow the treatment from Ref. \cite{Denner:2005fg} and write the renormalized counterterms as 
\begin{gather}
\delta \mu_W^2=\Sigma_T^W(M_W^2)+(\mu_W^2-M_W^2)\Sigma^{\prime W}_T(M_W^2)+c_T^W+\mathcal{O}(\alpha^3),\nonumber\\
\delta \mu_Z^2=\Sigma_T^{ZZ}(M_Z^2)+(\mu_Z^2-M_Z^2)\Sigma^{\prime Z}_T(M_Z^2)+\mathcal{O}(\alpha^3),\nonumber\\
\delta \mathcal{Z}_W=-\Sigma^{\prime W}_T(M_W^2),\quad \delta \mathcal{Z}_{ZZ}=-\Sigma^{\prime ZZ}_T(M_Z^2),\nonumber\\
\delta \mathcal{Z}_H =-\Sigma^{\prime H}(M_H^2),\quad \delta \mathcal{Z}_{ZA}=\frac{2}{\mu_Z^2}\Sigma_T^{AZ}(0),\quad \delta \mathcal{Z}_{AA}=-\Sigma^{\prime AA}_T(0).
\label{eq:ct_terms_cms_cal}
\end{gather}
$\delta Z_e$ follows same equation as given in Eq.~\ref{eq:ct_terms_cal}. In Eq.~\ref{eq:ct_terms_cms_cal}, the 'Re' part is taken out in contrast to the Eq.~\ref{eq:ct_terms_cal} as the self energies become complex in CMS via their complex masses and couplings. The extra term $c_T^W$($=\frac{i\alpha}{\pi}M_W\Gamma_W$) in $\delta \mu_W^2$ comes from the extra photon exchange diagram in $W$-boson self energy that has the branch cut at $k^2=\mu^2_W$. With the counterterms given in Eq.~\ref{eq:ct_terms_cms_cal}, we do the renormalization for this process and remove all UV divergences from the one-loop virtual amplitudes in CMS.

\subsection{Input parameter scheme}
\label{subsec:input_ps}
  The input parameters for EW corrections should be taken in a consistent way to get the right results. As we do on-shell renormalization, the pole masses of massive fermions and vector bosons have been used in our computation.
  The Weinberg angle is not an independent parameter, but written in terms of $W$ and $Z$ boson masses. The convenient choice for input parameters for EW correction are the electromagnetic coupling $\alpha$, masses of vector bosons $M_Z$ and $M_W$, Higgs boson mass $M_H$ and the fermion masses. Depending on the choice of the scale of the process, the coupling $\alpha$ may differ by a few percent, so the choice of the weak coupling also has an impact on results.

  The charge renormalization constant $\delta Z_e$ is calculated from the photon self-energy renormalization constant $\delta Z_{AA}$ as it can be seen from Eq.~\ref{eq:ct_terms_cal}. 
  The renormalization constant $\delta Z_{AA}$ contains mass singular terms $\alpha \log m_f$ where $m_f$ is the mass of the fermion. These contribution comes from every light fermion loop in $\delta Z_{AA}$ and remains uncancelled in EW corrections. 
  To renormalize the electric charge, the standard QED on-shell renormalization condition is being imposed in the Thomson limit , where the photon momentum transfer is zero. This renormalizes the QED coupling $\alpha=\alpha(0)$ at $Q^2=0$. To have the weak coupling $\alpha$ at desire scale ($Q^2\sim M^2_Z$) one needs running of $\alpha$ from $Q^2=0$ to $Q^2=M^2_Z$. The running of $\alpha$ remove the mass singular terms from the charge renormalization. 
  
  The choice of the running of the coupling $\alpha$ leads to the notion of the input parameter scheme. In $\alpha(M_Z)$ input parameter scheme, the $\Delta \alpha(M_Z)$ is given by~\cite{Andersen:2014efa}
  \begin{gather}
    \Delta\alpha(M_Z)=\frac{\alpha(0)}{3\pi}\sum_{f\neq t}N_f^cQ^2_f\Big[\text{ln}\Big(\frac{M_Z^2}{m_f^2}\Big)-\frac{5}{3}\Big].
   \label{eq:da_amzs}
  \end{gather}
  The shift in charge renormalization $\delta Z_e |_{\alpha(M_Z^2)}\rightarrow \delta Z_e|_{\alpha(0)}-\frac{1}{2}\Delta \alpha(M_Z^2)$, will remove all mass singularities in $\delta Z_e$. The numerical value of $\alpha(M_Z)$ has been extracted from an experimental analysis of $e^+e^-$ annihilation to hadrons~\cite{Eidelman:1995ny}.
  In $\alpha_{G_F}$ scheme, the electromagnetic coupling is derived from the Fermi constant as 
  \begin{gather}
    \alpha_{G_F}=\frac{\sqrt{2}G_FM_W^2(M_Z^2-M_W^2)}{\pi M_Z^2}.
   \label{equ_h24nu_alp_gms}
  \end{gather}
  In this scheme, the shift in charge renormalization is given by $\delta Z_e |_{G_F}\rightarrow \delta Z_e|_{\alpha(0)}-\frac{1}{2}\Delta r$, where the $\Delta r$ is the radiative correction to muon decay~\cite{PhysRevD.22.971,Denner:1991kt}. The $\Delta r$ is given by
  \begin{eqnarray}
    \Delta r &=& \Sigma^{AA}_T(0)-\frac{c_W^2}{s_W^2}\Big(\frac{\Sigma_T^{ZZ}(M_Z^2)}{M_Z^2}-\frac{\Sigma_T^W(M_W^2)}{M_W^2}\Big)+\frac{\Sigma^W_T(0)-\Sigma^W_T(M_W^2)}{M_W^2}\nonumber \\
    &&\quad \quad +2\:\frac{c_W}{s_W}\frac{\Sigma^{AZ}_T(0)}{M_Z^2}+\frac{\alpha}{4\pi s_W^2}\Big(6+\frac{7-4s_W^2}{2s_W^2}{\text {log}}\:c_W^2\Big).
   \label{eq:dr_gms}
   \end{eqnarray}
   We calculate EW correction to this process in the both $\alpha(M_Z)$ and $G_F$ schemes. The ``best" scheme will be the one in which the universal correction will be absorbed into the corresponding lower order prediction and leading to smaller perturbative correction.
   We will see in Sec.~\ref{sec_h24nu_numr_res}, the EW correction is smaller in the $G_F$ scheme than $\alpha(M_Z)$ scheme. Hence, the $G_F$ scheme can be regarded as the ``best" input scheme for this process.
\color{black}
\subsection{Anomalous couplings}
\label{subsec:renorm_cms}
As discussed in the above sections, our main goal is to study the  effect of anomalous $HHH$, $ZZHH$ and, $ZZWW$ couplings in this process. There are a few virtual diagrams as displayed in Fig.~\ref{fig:hhh_dia}, Fig.~\ref{fig:zzhh_dia}, and  Fig.~\ref{fig:wwzz_dia}, where we can introduce such anomalous couplings in the $\kappa$-framework. The $HHH$ coupling is involved in the diagrams shown in the Fig.~\ref{fig:hhh_dia} and in the Higgs boson wave function renormalization constant which has been computed from its self-energy diagrams. The sum of the triangle diagrams shown in Fig.~\ref{fig:hhh_dia} is UV finite and the contribution of the related diagrams to the Higgs boson wave function renormalization constant is also UV finite. With this UV pole structure, the renormalizability is sustained even after arbitrary scaling of $HHH$ coupling in this process. The reason behind this UV structure is that the $HHH$ coupling comes from the potential term of the standard model Lagrangian and it does not get coupled with the other terms in the Lagrangian. Therefore, we can vary the $HHH$ coupling within the allowed region and study its effect on the partial decay width of the Higgs boson.

The $ZZHH$ and $ZZWW$ couplings can be scaled in the virtual diagrams displayed in Fig.~\ref{fig:zzhh_dia} and Fig.~\ref{fig:wwzz_dia}. There are also other places to vary these couplings in various counterterms that involve $W$, $Z$ and, Higgs boson self energies. Scaling these two couplings in the diagrams lead to renormalization problem within SM as $HVV$ and $VVHH$ ($V=Z,W$) couplings are not independent. In this regard, we recall that in the HEFT framework, 
one can vary $HVV$ and $VVHH$ couplings independently. Furthermore, many beyond-the-standard-model scenarios have different relationship
between HVV and HHVV couplings than that in the standard model. The excess UV pole contribution arising from the scaling of the couplings can be absorbed in the $HZZ$ coupling, as, for example, in the context of HEFT. We adopt $\overline{MS}$ scheme for scaled couplings. Therefore, we only cancel excess UV divergent piece with appropriate
counterterm. Then, we vary $ZZHH$ and $ZZWW$ couplings and study their effect on the Higgs decay width.

\section{Numerical Results}

\label{sec_h24nu_numr_res}
\subsection{SM prediction}
\label{subsec_h24nu_sm_pdct}
We use the following set of the standard model input parameters
\begin{gather}
 M_W=80.358\: {\text{GeV}},\:\:M_Z=91.153\: {\text{GeV}},\:\:M_t=172.5\: {\text{GeV}}\:\:M_H=125\: {\text{GeV}},\nonumber\\
\Gamma_W=2.0872 \: {\text{GeV}},\:\:\Gamma_Z=2.4944 \: {\text{GeV}},\:\Gamma_H=4.187\: {\text{MeV}} \:{\text{and}}\: \Gamma_t=1.481 \: {\text{GeV}}.
\label{eq_h24nu_sm_param}
\end{gather}
We have taken lepton and light quarks as massless particles.
    Mass parameters in Eq.~\ref{eq_h24nu_sm_param} and cosine of Weinberg angle have been promoted to complex numbers following the Eq.~\ref{eq:cms_equ_cal}.
We calculate this process in the both $\alpha(M_Z)$ and $G_F$ input parameter schemes. The value of electromagnetic coupling $\alpha=1/128.896$ for the $\alpha(M_Z)$ scheme has been taken from the Ref.~\cite{Eidelman:1995ny}.
   In the $G_F$ scheme, the value of electromagnetic coupling can be calculated from the Eq.~\ref{equ_h24nu_alp_gms}. With the above SM parameters, its numerical value is $1/132.36$ in the $G_F$ scheme. 
   The standard model prediction for the partial decay width of Higgs boson for this process has been listed in Tab.~\ref{tab_h24nu_sm_rst}.
   The  LO decay widths are $930.71$ eV and $1007.72$ eV in the $G_F$ and $\alpha(M_Z)$ scheme respectively. The NLO corrected decay widths are $959.66$ eV and $948.01$ eV in the $G_F$ and $\alpha(M_Z)$ schemes respectively. We define the relative enhancement as ${\text{RE}}=\frac{\Gamma^{NLO}-\Gamma^{LO}}{\Gamma^{LO}}\; \times 100\%$.
   The RE is $3.11\%$ in the $G_F$ scheme whereas it is $-5.93\%$ in the $\alpha(M_Z)$ scheme. The LO decay widths differ by $\sim 8\%$ but the NLO corrected widths differ by $\sim 1\%$ among the two input parameter schemes.
     As we can see from Tab.~\ref{tab_h24nu_sm_rst}, the relative enhancement is smaller in the $G_F$ scheme i.e., the universal correction has been absorbed in the lower order prediction, so, this scheme can be considered as the ``better" scheme.

\begin{table}[H]
\begin{center}
\begin{tabular}{|c|c|c|c|}
\hline
{Input}&&&\\
{parameter}&{$\Gamma^{LO}$ (eV)}&{$\Gamma^{NLO}$ (eV)}&{RE}\\
{scheme}&&&\\
\hline
&&&\\
{$G_F$}&{$930.71$}&{$959.66$}&{$3.11\%$}\\
&&&\\

\hline
&&&\\
{$\alpha(M_Z)$}&{$1007.72$}&{$948.01$}&{$-5.93\%$}\\
&&&\\
\hline
\end{tabular}
\caption{{Partial decay widths of Higgs boson in the channel $H\rightarrow \nu_e \bar{\nu}_e \nu_\mu \bar{\nu}_\mu$ in the $G_F$ and $\alpha(M_Z)$ scheme and their relative enhancement.}}
\label{tab_h24nu_sm_rst}
\end{center}
\end{table}
\subsection{Anomalous coupling effect}
\label{subsec_h34nu_anoma_cpl_efct}
  We vary $HHH$, $ZZHH$ and $ZZWW$ couplings in the context of kappa ($\kappa$) framework. We define relative increment as ${\text{RI}}=\frac{\Gamma^{NLO}_\kappa-\Gamma^{NLO}_{SM}}{\Gamma^{NLO}_{SM}}\; \times 100\%$.

  We have varied $\kappa_{HHH}$ from $-10$ to $10$ and listed the corresponding RI in Tab.~\ref{tab_h24nu_hhh_kappa}. In Tab.~\ref{tab_h24nu_hhh_kappa}, we see a significant change in NLO EW decay width ($\Gamma^{NLO}$) with varying $\kappa_{HHH}$.
  The RI varies from $\sim 0.35\%$ to $\sim -23.52\%$ in the $G_F$ scheme and from $\sim 0.40\%$ to $\sim -26.48\%$ in the $\alpha(M_Z)$ scheme depending on
the value of $\kappa_{HHH}$. The RI in two input schemes become positive near $k_{HHH}\sim 2{\text{-}}4$.
\begin{table}[H]
\begin{center}
\begin{tabular}{|c|c|c|}
\hline
\multirow{2}{*}{{$\kappa_{HHH}$}}&\multicolumn{2}{|c|}{{RI}}\\
\cline{2-3}
&{$G_F$ scheme}&{$\alpha(M_Z)$ scheme}\\
\hline
{$10$}&{$-7.54$}&{$-8.48$}\\
\hline
{$8$}&{$-3.78$}&{$-4.25$}\\
\hline
{$6$}&{$-1.21$}&{$-1.36$}\\
\hline
{$4$}&{$0.17$}&{$0.19$}\\
\hline
{$2$}&{$0.35$}&{$0.40$}\\
\hline
{$-1$}&{$-1.60$}&{$-1.80$}\\
\hline
{$-2$}&{$-2.84$}&{$-3.20$}\\
\hline
{$-4$}&{$-6.23$}&{$-7.01$}\\
\hline
{$-6$}&{$-10.80$}&{$-12.16$}\\
\hline
{$-8$}&{$-16.57$}&{$-18.65$}\\
\hline
{$-10$}&{$-23.52$}&{$-26.48$}\\
\hline
\end{tabular}
\caption{{Effect of anomalous $HHH$ coupling on the partial decay width of the process $H\rightarrow\nu_e \bar{\nu}_e \nu_\mu \bar{\nu}_\mu$.}}
\label{tab_h24nu_hhh_kappa}
\end{center}
\end{table}
%
%

Next, we examine the effect of scaling the $ZZHH$ coupling. We have listed the RI with different $\kappa_{ZZHH}$ values in Tab.~\ref{tab_h24nu_zzhh_kappa}. As shown in the table, the change in RI due to $\kappa_{ZZHH}$ is hardly visible in the $\alpha(M_Z)$ scheme. It is less than $1 \%$.
 In the $G_F$ scheme, the RI varies from $\sim 5.7\%$ to $\sim -7.0\%$ depending upon $\kappa_{ZZHH}$.
It is positive with positive scaling and negative with negative scaling.
\begin{table}[H]
\begin{center}
\begin{tabular}{|c|c|c|}
\hline
\multirow{2}{*}{{$\kappa_{ZZHH}$}}&\multicolumn{2}{|c|}{{RI}}\\
\cline{2-3}
&{$G_F$ scheme}&{$\alpha(M_Z)$ scheme}\\
\hline
{$10$}&{$5.74$}&{$0.29$}\\
\hline
{$8$}&{$4.46$}&{$0.22$}\\
\hline
{$6$}&{$3.19$}&{$0.16$}\\
\hline
{$4$}&{$1.91$}&{$0.10$}\\
\hline
{$2$}&{$0.64$}&{$0.03$}\\
\hline
{$-1$}&{$-1.27$}&{$-0.06$}\\
\hline
{$-2$}&{$-1.91$}&{$-0.09$}\\
\hline
{$-4$}&{$-3.19$}&{$-0.16$}\\
\hline
{$-6$}&{$-4.46$}&{$-0.22$}\\
\hline
{$-8$}&{$-5.74$}&{$-0.29$}\\
\hline
{$-10$}&{$-7.01$}&{$-0.35$}\\
\hline
\end{tabular}
\caption{{Effect of anomalous $ZZHH$ coupling on the partial decay width of the process $H\rightarrow \nu_e \bar{\nu}_e \nu_\mu \bar{\nu}_\mu$.}}
\label{tab_h24nu_zzhh_kappa}
\end{center}
\end{table}
%

  Although it was not our main goal, since the width of the process
  also depends on the $ZZWW$ coupling, we scale it to see the effects.
We have listed the RI by scaling $ZZWW$ coupling in Tab.~\ref{tab_h24nu_zzww_kappa}. As shown in Tab.~\ref{tab_h24nu_zzww_kappa}, we see a significant change in $\Gamma^{NLO}$ with varying $\kappa_{ZZWW}$.
 In the $G_F$ scheme, the RI goes from $\sim 10.5$ to $\sim -12.8\%$, whereas in the $\alpha(M_Z)$ scheme  it goes from $-19.3\%$ to $23.5\%$ with the varying $\kappa_{ZZWW}$ from $10$ to $-10$.
\begin{table}[H]
\begin{center}
\begin{tabular}{|c|c|c|}
\hline
\multirow{2}{*}{{$\kappa_{ZZWW}$}}&\multicolumn{2}{|c|}{{RI}}\\
\cline{2-3}
&{$G_F$ scheme}&{$\alpha(M_Z)$ scheme}\\
\hline
{$10$}&{$10.45$}&{$-19.29$}\\
\hline
{$8$}&{$8.13$}&{$-14.97$}\\
\hline
{$6$}&{$5.81$}&{$-10.72$}\\
\hline
{$4$}&{$3.48$}&{$-6.43$}\\
\hline
{$2$}&{$1.16$}&{$-2.14$}\\
\hline
{$-1$}&{$-2.32$}&{$4.29$}\\
\hline
{$-2$}&{$-3.48$}&{$6.43$}\\
\hline
{$-4$}&{$-5.80$}&{$10.72$}\\
\hline
{$-6$}&{$-8.13$}&{$15.00$}\\
\hline
{$-8$}&{$-10.45$}&{$19.29$}\\
\hline
{$-10$}&{$-12.78$}&{$23.58$}\\
\hline
\end{tabular}
\caption{{Effect of anomalous $ZZWW$ coupling on the partial decay width of the process $H\rightarrow\nu_e \bar{\nu}_e \nu_\mu \bar{\nu}_\mu$.}}
\label{tab_h24nu_zzww_kappa}
\end{center}
\end{table}
%
%
%

\section{Conclusion}
\label{sec_h24nu_conclusion}

     We have studied the effect of scaling $HHH$ and $ZZHH$
     couplings, as in $\kappa$-framework, on the decay width of the
     process $H\rightarrow \nu_e\bar{\nu}_e\nu_\mu\bar{\nu}_\mu$.
     As this process also depends on $ZZWW$ coupling, we also investigated
     the effect of its variation. Most interesting thing we find is
     that the width of this process has significant dependence on the
     $\kappa_{HHH}$. 
    The dependency on $\kappa_{HHH}$ is similar among the two input parameter scheme.
    As the scaling of $HHH$ coupling does not effect the gauge invariance, we see similar behaviour in the two input schemes. 
     A precise measurement of the width may lead to a better bound on the $HHH$ coupling.
     We also examine the dependency of $\kappa_{ZZHH}$ on decay partial width.
      We see very minimal dependency on $\kappa_{ZZHH}$ for the $\alpha(M_Z)$ scheme, but in the $G_F$ scheme, we see a bit stronger dependency on $\kappa_{ZZHH}$. 
      This is due to the terms $\alpha(M_Z)$ and $\Delta r$ in charge renormalization $\delta Z_e$ in two input parameter schemes. In the $\alpha(M_Z)$ in Eq.~\ref{eq:da_amzs}, there is no $\Sigma^{ZZ}$ self-energy term, so no dependency on $\kappa_{ZZHH}$; whereas in the term $\Delta r$ in Eq.~\ref{eq:dr_gms}, there is $\Sigma^{ZZ}$ self-energy term, hence dependency on $\kappa_{ZZHH}$.
      Because of the dependence of
     the process on $\kappa_{ZZWW}$, we also examined the dependence
     on this parameter. However, there are other processes to better
     determine this coupling.

\color{black}
 \section*{Acknowledgements}
     We would like to acknowledge useful discussions with C.-P. Yuan and
     Ambresh Shivaji.  
\input{h22vm2ve.bbl}
\end{document}

%% file: h22vm2ve.bbl
\providecommand{\href}[2]{#2}\begingroup\raggedright\endgroup